\documentclass[preprint2]{aastex}

\usepackage{graphicx}
\usepackage{subfigure}

\shorttitle{On the average density profile of dark-matter halos \\
in the inner regions of massive early-type galaxies}

\begin{document}


\title{On the average density profile of dark-matter halos \\
in the inner regions of massive early-type galaxies}


\author{C. Grillo\altaffilmark{1}}
\email{grillo.claudio@googlemail.com}


\altaffiltext{1}{Excellence Cluster Universe, Technische Universit\"at M\"unchen, Boltzmannstr. 2, D-85748, Garching bei M\"unchen, Germany}


\begin{abstract}

We study a sample of 39 massive early-type lens galaxies at redshift $z \lesssim 0.3$ to determine the slope of the average dark-matter density profile in the innermost regions. We keep the strong lensing and stellar population synthesis modeling as simple as possible to measure the galaxy total and luminous masses. By rescaling the values of the Einstein radius and dark-matter projected mass with the values of the luminous effective radius and mass, we combine all the data of the galaxies in the sample. We find that between 0.3 and 0.9 times the value of the effective radius the average logarithmic slope of the dark-matter projected density profile is $-1.0 \pm 0.2$ (i.e., approximately isothermal) or $-0.7 \pm 0.5$ (i.e., shallower than isothermal), if, respectively, a constant Chabrier or heavier, Salpeter-like stellar IMF is adopted. These results provide positive evidence of the influence of the baryonic component on the contraction of the galaxy dark-matter halos, compared to the predictions of dark matter-only cosmological simulations, and open a new way to test models of structure formation and evolution within the standard $\Lambda$CDM cosmological scenario.
 
\end{abstract}


\keywords{galaxies: elliptical and lenticular, cD $-$ galaxies: structure $-$ dark matter}



\section{Introduction}

Understanding the original formation and the subsequent evolution of the galaxies we observe today remains one of the major open questions in modern astrophysics. Within the past decade, there has been incredible progress in the realization of high-resolution numerical simulations that are now starting to reproduce in detail the physical properties of the galaxies observed at redshift $z = 0$ (e.g., \citealt{mez03}; \citealt{naa07}; \citealt{tis10}). $N$-body cosmological simulations have predicted that in an expanding universe cold dark-matter particles collapse into gravitationally bound, self-similar halos with a diverging inner density profile (e.g., \citealt{nav96}; \citealt{moo99}). The values of the three-dimensional logarithmic slope $\gamma = \mathrm{d} \ln \rho / \mathrm{d} \ln r$ of the collapsed dark-matter halos have been found to be approximately equal to $-1$ and $-3$, respectively, in the innermost and outermost regions. It is in these halos that the stars of the observed galaxies were assembled. Recent hydrodynamical simulations have shown that several mechanisms associated to baryonic physics affect the stellar mass assembly of a galaxy (e.g., dissipationless accretion of stars originally formed far from a galaxy center and dissipational gas flowing towards the inner regions of a galaxy, later transformed into stars). The complex interplay between the luminous and dark components can alter significantly the dark-matter distribution in the center of a halo, making it steeper or shallower, depending on the role played by the different physical processes (for more details, see e.g. \citealt{lac10}).

In the last two decades strong gravitational lensing combined with stellar dynamics and/or stellar population synthesis models has been extremely successful in measuring the amount and distribution of dark matter (\citealt{gri08c,gri10a,gri11}; \citealt{bar07,bar09}; \citealt{aug09,aug10a}; \citealt{fad10}), the presence of dark-matter substructure (e.g., \citealt{veg09,veg10}), and the sizes of dark-matter halos (e.g., \citealt{suy10a}; \citealt{ric10}; \citealt{don11}) in early-type galaxies beyond the local Universe. The combination of these mass diagnostics has also allowed to find alternative ways to address some interesting astrophysical and cosmological topics, such as the determination of the stellar initial mass function (IMF) (e.g., \citealt{gri08a,gri09,gri10b}; \citealt{tre10}; \citealt{aug10b}; \citealt{spi11}; \citealt{son11}) and of the values of the cosmological parameters (e.g., \citealt{gri08b}; \citealt{par09}; \citealt{sch10}; \citealt{suy10b}).

The Sloan Lens ACS (SLACS) survey has been crucial for the identification of a statistically significant sample of strong gravitational lensing systems. Disparate studies (e.g., \citealt{tre06}; \citealt{bol06}; \citealt{gri09}; \citealt{aug09}) have shown that the SLACS lens galaxies are a representative sample of the parent sample of massive early-type galaxies observed in the Sloan Digital Sky Survey (SDSS). By modeling the strong gravitational features detected in these lensing systems, it has been possible to obtain accurate and precise total mass estimates projected within the corresponding Einstein radii (\citealt{tre06}; \citealt{koo06}; \citealt{bol08a}; \citealt{aug09}). Here, we exploit the fact that the Einstein radius of a lensing system is not a length scale intrinsic to the lens (since it depends also on the redshift of the source) to study the average inner dark-matter density distribution of a specific lens sample. We do this by combining the lens aperture total and luminous mass measurements. The pioneering work of \citet{rus03} prefigures to some extent the general method and results presented here. In this previous analysis a self-similar mass model for early-type galaxies was constrained by using aperture mass-radius relations from 22 gravitational lenses. The total mass distribution of the lens galaxies was described in terms of a two-component (luminous and dark matter) model parametrized by (1) a present-day normalization value of the $B$-band stellar mass-to-light ratio, (2) the dependence of a galaxy $B$-band stellar mass-to-light ratio on its luminosity, (3) the projected dark over total mass fraction within two effective radii and (4) the three-dimensional logarithmic density slope of the dark-matter profile.

This Letter is organized as follows. In Sect. 2, we introduce the sample of massive early-type lens galaxies. In Sect. 3, we describe the method and hypotheses used to determine the inner slope of the average galaxy dark-matter density profile. In Sect. 4, we illustrate the main results of this analysis. In Sect. 5, we compare our results with those of previous studies and anticipate future prospects. In Sec. 6, we draw conclusions. In the following, we assume $H_{0}=70$ km s$^{-1}$ Mpc$^{-1}$, $\Omega_{m}=0.3$, and $\Omega_{\Lambda}=0.7$. 

\section{The sample}

\begin{table*}
\centering
\caption{Physical properties of the early-type lens galaxies of the sample.}
\begin{tabular}{cccccc}
\hline\hline \noalign{\smallskip}
$z_{\mathrm{sp}}$ & $R_{e}$ & $R_{\mathrm{Ein}}$ & $\sigma_{0}$ & $M_{L}$ & $M_{T}(<R_{\mathrm{Ein}})$ \\
 & (kpc) & (kpc) & (km s$^{-1}$) & ($10^{10}$ $M_{\odot}$) & ($10^{10}$ $M_{\odot})$ \\
\noalign{\smallskip} \hline
0.06-0.32 & 3.2-16 & 1.3-7.0 & 200-320 & 7.4-56 & 3.9-47 \\
\noalign{\smallskip} \hline
\end{tabular}
\begin{list}{}{}
\item[Notes --]Ranges of values of the spectroscopic redshift $z_{\mathrm{sp}}$, effective radius $R_{e}$, Einstein radius $R_{\mathrm{Ein}}$, central stellar velocity dispersion $\sigma_{0}$, luminous mass $M_{L}$ (assuming a constant Chabrier stellar IMF), and total mass projected within the Einstein radius $M_{T}(<R_{\mathrm{Ein}})$.
\item[References --]SDSS and MPA/JHU public catalogs; \citet{aug09}; \citet{gri10c}.
\end{list}
\label{ta02}
\end{table*}

In this work, we concentrate on 39 massive early-type lens galaxies discovered in the SLACS survey and studied in several papers (e.g., \citealt{bol08a}; \citealt{gri09}; \citealt{aug10a}). In detail, we conservatively consider only those galaxies that satisfy the photometric and spectroscopic selection criteria of the sample analyzed in \citet{gri10c} (i.e., values of the SDSS \textsf{fracDeV} morphological index larger than 0.95 in the $r$, $i$, and $z$ bands; SDSS spectroscopic redshifts $z_{\mathrm{sp}}$ between 0.05 and 0.33; SDSS aperture stellar velocity dispersions between 150 and 400 km s$^{-1}$; total luminous masses between $10^{10.5}$ ans $10^{12}$ $M_{\odot}$). These galaxies have both accurate total $M_{T}$ and luminous $M_{L}$  mass estimates obtained from, respectively, strong lensing (\citealt{aug09}) and spectral energy distribution (SED) fitting models (from the public galaxy catalogs provided by the MPA/JHU collaboration\footnote{http://www.mpa-garching.mpg.de/SDSS/}). The physical properties of the galaxies in the sample are summarized in Table \ref{ta02}. This is a specific sample of early-type galaxies with large values of central stellar velocity dispersion $\sigma_{0}$ (more details on the measurements of the physical quantities can be found in \citealt{gri10c}). Therefore, the results of the analysis performed in this letter should not be simplistically generalized to early-type galaxies with different physical properties until verified by larger samples.

\section{The method}

For the lenses in the sample, we measure here the values of the dark-matter mass density projected within the Einstein radius and study, in a statistical way, their dependence on the projected distance from the lens centers.

In practice, we proceed as follows. For each lens galaxy, we define an adimensional radius $\Lambda$ as the ratio between the Einstein radius $R_{\mathrm{Ein}}$ and the effective radius $R_{e}$:  
\begin{equation}
\Lambda := \frac{R_{\mathrm{Ein}}}{R_{e}} \, .
\label{eq:01}
\end{equation}
The Einstein radius of a lens galaxy depends on its total mass distribution, but also on the redshift of the lensed source. Thus, $R_{\mathrm{Ein}}$ is not a fundamental property of a galaxy and we use $\Lambda$ instead to quantify the distance from the center of a lens. The latter is a scale-free distance that is obtained by normalizing the value of the Einstein radius to the typical scale of a galaxy luminous mass distribution (i.e., $R_{e}$). 

Similarly, we estimate the value of the dark-matter mass projected inside the cylinder with radius equal to the Einstein radius as the difference between the values of the total $M_T(<R_{\mathrm{Ein}})$ and luminous $M_L(<R_{\mathrm{Ein}})$ masses and rescale the result to the total amount of luminous mass $M_L$ of each galaxy. We notice that the measurements of the projected masses within $R_{\mathrm{Ein}}$ are robust and almost model-independent for the total ones and only scaled according to the fraction of total light of a de Vaucouleurs profile for the luminous ones (see \citealt{gri09,gri10c}). Then, we define an adimensional dark-matter projected mass density $\Psi$ as the ratio between the adimensional value of the dark-matter projected mass and the area of the disk with radius equal to the value of $\Lambda$:
\begin{equation}
\Psi := \frac{M_T(<R_{\mathrm{Ein}})-M_L(<R_{\mathrm{Ein}})}{M_L} \frac{1}{\pi \Lambda^{2}} \, .
\label{eq:02}
\end{equation}
In this way, both $\Lambda$ and $\Psi$ are referred to the luminous properties of the galaxies in the sample and can thus be properly compared.\footnote{In passing, we notice that differently from \citet{rus03} the luminous mass values of the lens galaxies are measured here from the multi-band photometric and spectroscopic observables and are not scaled according to the galaxy $B$-band luminosity values.}

Next, we measure the value of the Kendall rank correlation coefficient $\varrho$ (for its definition, see \citealt{sal06}) and check if the values of $\Lambda$ and $\Psi$ are correlated at a statistically significant level. In the case of a significant correlation, we perform a Markov chain Monte Carlo study on the galaxy sample to characterize the joint probability distribution function of the values of the two coefficients $\alpha$ and $\beta$ that are used to fit a power-law relation to our set of data:
\begin{equation}
\Psi = \alpha \times (\Lambda)^{\beta} \, .
\label{eq:03}
\end{equation}

We apply this method to the sample described in Sect. 2, starting from different hypotheses. First, we assume a constant \citet{cha03} (labeled as Ch) stellar IMF to estimate the luminous mass values of all the galaxies in the sample. Then, we rescale the galaxy luminous mass values to a constant heavier, \citet{sal55}-like (labeled as Sa) stellar IMF by simply multiplying the Chabrier luminous mass values by a constant factor equal to 1.7. Next, we consider the case of a non-universal stellar IMF and mimic a variation, moving from a lighter to a heavier IMF (labeled as Ch $\rightarrow$ Sa), depending on the values of the galaxy central stellar velocity dispersion. This is motivated by the facts that stellar velocity dispersion is currently considered the most significant parameter related to the stellar population properties of a galaxy (e.g., \citealt{gra09}) and that a stellar IMF variation with stellar velocity dispersion has been tentatively detected by \citet{tre10}. In detail, following the previous indications, we use a toy model in which we multiply the Chabrier luminous mass values with a factor that increases linearly from 1.0 to 1.5 as the value of $\sigma_{0}$ changes from 200 to 320 km s$^{-1}$.

We conclude by remarking that the correlation of the errors on $\Lambda$ and $\Psi$ is not significant and therefore will not affect our results on the steepness of the average dark-matter density profile. Although obtained from the same sets of observational quantities, the uncertainties on $\Lambda$ are very small (the median relative error is smaller than 4\%) and mainly related to the quality of the photometric measurements, while the uncertainties on $\Psi$ are considerably large (the median relative error is approximately 40\%) and primarily driven by the degeneracies that are inherent in the population synthesis modeling.

\section{Results}

\begin{table*}
\centering
\caption{Correlations and power-law fits of $\Lambda$ and $\Psi$.}
\begin{tabular}{cccc}
\hline\hline \noalign{\smallskip}
 & $\varrho(\Lambda,\Psi)$ & $\beta_{\mathrm{best}}$ & $\beta_{68\%}$$\,$$_{\mathrm{CL}}$ \\
\noalign{\smallskip} \hline
Ch & $-0.57$ $(<0.01)$ & $-$1.04 & $[-1.26,-0.78]$ \\
Sa & $-0.24$ $(<0.03)$ & $-$0.77 & $[-1.14,-0.15]$ \\
Ch $\rightarrow$ Sa & $-0.52$ $(<0.01)$ & $-$1.28 & $[-1.54,-0.93]$ \\
\noalign{\smallskip} \hline
\end{tabular}
\begin{list}{}{}
\item[Notes --]Values of the Kendall rank correlation coefficient $\varrho$ between $\Lambda$ and $\Psi$ (in parentheses, the probability that an equal number of measurements of two uncorrelated variables would give values of the coefficient higher than the measured ones), and of the best-fitting $\beta_{\mathrm{best}}$ and 68\% CL interval $\beta_{68\%}$$\,$$_{\mathrm{CL}}$ of the inner slope of the average dark-matter projected mass density.
\end{list}
\label{ta01}
\end{table*}

\begin{figure}[tb]
\centering
\includegraphics[width=0.49\textwidth]{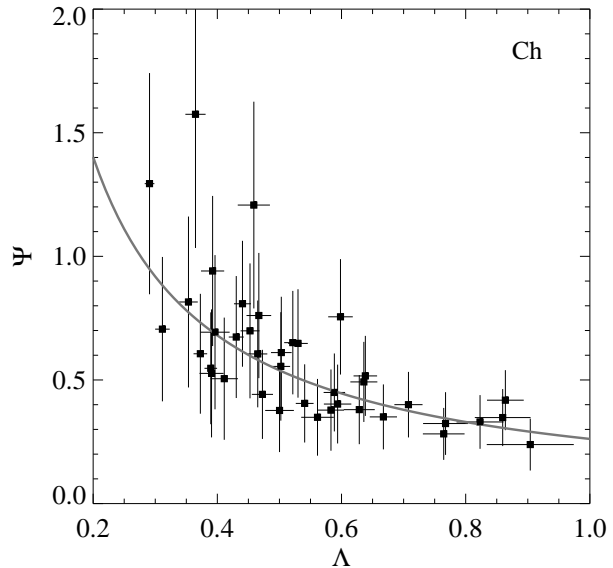}
\caption{The adimensional values of the dark-matter projected mass density within the Einstein radius, $\Psi$, and Einstein radius, $\Lambda$. The points, with their 1 $\sigma$ error bars, are obtained by using the values of the total luminous mass and effective radius of the galaxies as dimensional scales and assuming a constant Chabrier stellar IMF. The best-fitting power-law is shown in gray.}
\label{fi01}
\end{figure}

\begin{figure}[tb]
\centering
\includegraphics[width=0.49\textwidth]{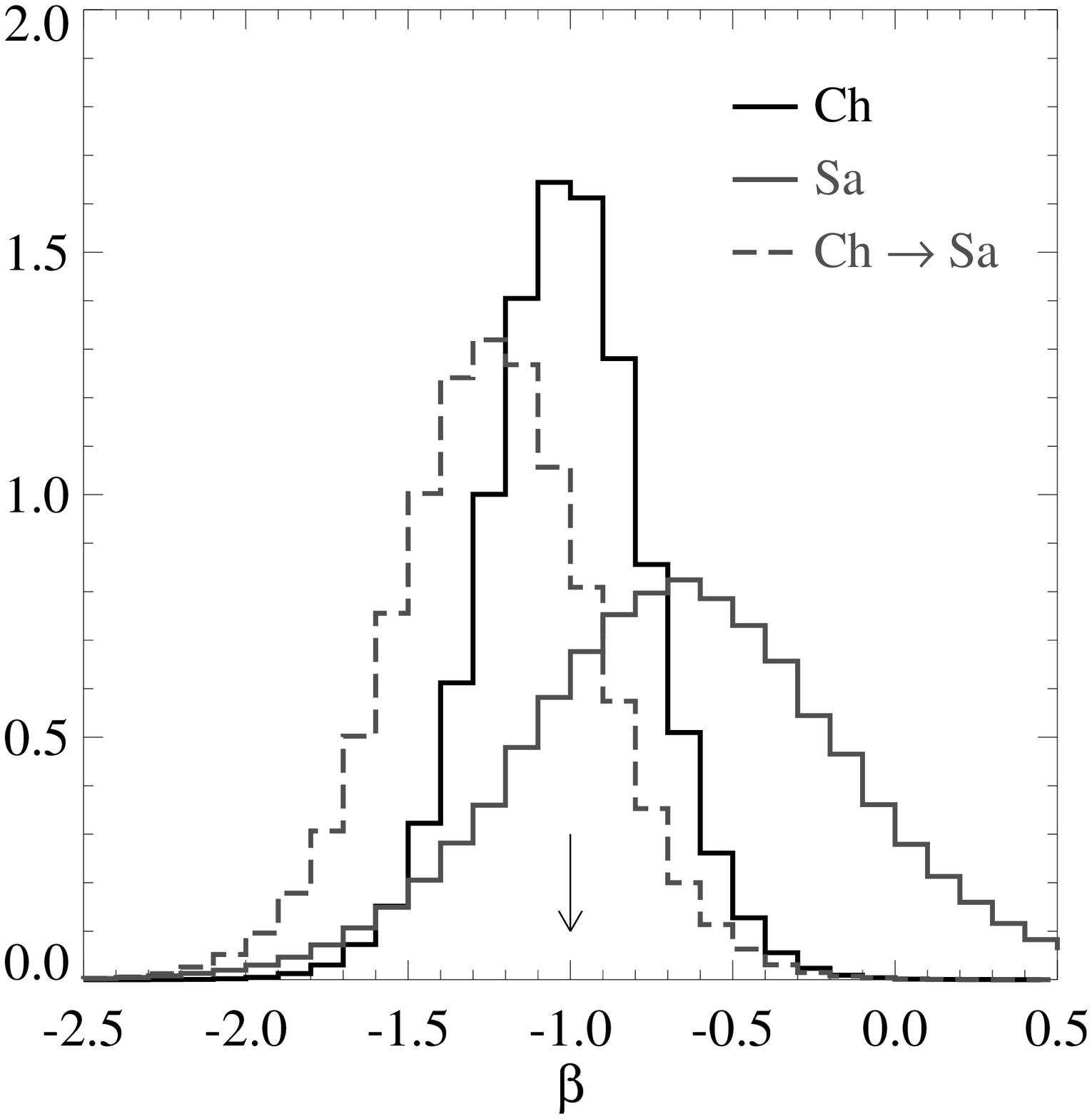}
\caption{Probability distribution functions of the average logarithmic inner slope, $\beta$, of the dark-matter projected mass density. The different histograms refer to the different assumptions on the stellar IMF discussed in the text. As a reference, the arrow close to the $x$-axis shows the result that a three-dimensional spherical density distribution decreasing as $1/r^{2}$ (i.e., an isothermal profile) would give. Larger and smaller values of $\beta$ correspond, respectively, to shallower and steeper profiles with respect to an isothermal one.}
\label{fi02}
\end{figure}

We summarize in Table \ref{ta01} the values of the Kendall rank correlation coefficient $\varrho$ between $\Lambda$ and $\Psi$ and remark that using the three different hypotheses mentioned above about the stellar IMF of the sample galaxies always results in an anti-correlation of the values of $\Lambda$ and $\Psi$ at a statistical significance level higher than 97\%. In the same table, we also show the best-fitting (minimum chi-square) $\beta_{\mathrm{best}}$ and the 68\% CL interval $\beta_{68\%}$$\,$$_{\mathrm{CL}}$ values of the average inner slope of the dark-matter projected mass density. These numbers are obtained from Monte Carlo chains with $5 \times 10^{5}$ points for each of the three cases. The data set and the best-fitting power-law for the case of a constant Chabrier stellar IMF are illustrated in Fig. \ref{fi01} and the marginalized probability distribution functions of $\beta$ for the three cases are plotted in Fig. \ref{fi02}.

From Table \ref{ta01} and Figs. \ref{fi01} and \ref{fi02}, we notice that assuming a constant Chabrier stellar IMF leads to an average dark-matter density profile that considered in terms of a three-dimensional spherical profile decreases in the inner regions approximately as $1/r^{2}$, i.e. like an isothermal profile (as usually referred to in lensing studies). A constant heavier stellar IMF results in a broader probability distribution function for $\beta$, centered on a slightly larger value. This result can be qualitatively explained in the following way. If we keep the values of the total mass fixed and increase those of the luminous mass, we obtain values of $\Psi$ that are on average smaller and decrease less steeply with increasing values of $\Lambda$ than in the previous case (see Fig. \ref{fi01}). This translates into a dark-matter density profile that is shallower than an isothermal one in the center. On the contrary, the proposed variation in the stellar IMF provides a steeper profile of the dark-matter component in the inner regions. This result can also be understood looking at Fig. \ref{fi01}. As expected, the values of $\sigma_{0}$ are positively correlated with those of $\Lambda$. This follows from the fact that more massive galaxies yield, on average, larger Einstein radii. Therefore, varying the stellar IMF from a Chabrier to a Salpeter-like, the points in Fig. \ref{fi01} with small values of $\Lambda$ have approximately the same values of $\Psi$ (because of the unchanged Chabrier stellar IMF), while those with large values of $\Lambda$ have now larger luminous mass values (because of the changed, heavier stellar IMF), hence, in general smaller values of $\Psi$. The net effect is an increase in the value of the slope $\beta$.    

We notice that the dark-matter universal profile obtained from dark matter-only cosmological simulations (\citealt{nav96}) is characterized by values of $\beta$ of approximately $-0.1$ and $-0.2$ within 0.1\% and 1\% the value of the typical dark-matter length scale ($r_{s}$), respectively.

\section{Discussion}

We compare here our results with those of several other studies on early-type galaxies and indicate a possible way to extend this work.

Based on a sample of 16 massive Coma galaxies, with physical properties very similar to those of the galaxies in our sample, \citet{tho11} find that if the stellar IMF is universal and \citet{kro01}-like, i.e. very similar to a Chabrier IMF, then the galaxy dark-matter density profiles are smooth and on average close to isothermal out to several tens of kiloparsecs (see Fig. 6 in the cited paper). This conclusion follows from joint dynamical (Schwarzschild's orbit superposition) and stellar population models that exploit accurate photometric and spectroscopic data. Recalling that the luminous mass estimates obtained by assuming a Chabrier or a Kroupa stellar IMF are only slightly different, the findings of our study on the value of $\beta$ in the 'Ch' case are consistent with those of the analysis performed in the Coma cluster.

\citet{nap10} consider a sample of 335 local early-type galaxies and estimate their luminous and total masses from, respectively, photometric SED fitting and dynamical Jeans modeling. They also adopt a Kroupa stellar IMF and conclude that the average three-dimensional logarithmic slope $\gamma$ of the dark-matter density profile at $R_{e}$ ranges between $-2.1$ and $-1.7$. In the simplified case of a spherical power-law density profile, the values of $\beta$ and $\gamma$ are related in projection in the following way: $\beta = \gamma + 1$ (if $\gamma$ is different from $-1$). Our estimates of $\beta$ in the 'Ch' case are therefore consistent with the results of \citet{nap10}.  

In the gravitational lensing study by \citet{rus03}, detailed in Sect. 1, the authors come to the conclusion that it is not possible to measure precisely the slope of the dark-matter component because of the significant degeneracies between the parameters. Despite that, models with a dark-matter density profile that is approximately isothermal are generally preferred to models with shallower dark-matter density distributions (e.g., with $\gamma = -1$). Our results confirm these last findings.

In the last few years, some new observational constraints have been obtained on the stellar IMF of massive early-type galaxies. For this specific class of galaxies, if the IMF is constant, a Salpeter-like IMF is favored by the data (e.g., \citealt{gri08a,gri09,gri10b}; \citealt{tre10}; \citealt{aug10b}; \citealt{spi11}). In the case of a constant Salpeter IMF, our results indicate a dark-matter density profile that is shallower than an isothermal one in the central regions. Interestingly, the study of \citet{tre04} on the average inner power-law slope of the dark-matter halos of 5 early-type lens galaxies at $z_{\mathrm{sp}} \approx 0.5-1.0$ provides also indication of profiles shallower than isothermal ones. The slope values are robustly determined by combining gravitational lensing and stellar dynamics, with and without priors on the lens stellar mass-to-light ratios from the Fundamental Plane. More recently, \citet{son11} have also performed a two-component lensing and dynamics analysis to decompose the total mass distribution of the double Einstein ring gravitational lens in terms of a bulge of stars and a dark-matter halo. They find that a Salpeter IMF is preferred to a Chabrier IMF for the stellar component and that the value of the three-dimensional logarithmic inner slope $\gamma$ of the dark-matter halo is $-1.7 \pm 0.2$. Therefore, our findings in the 'Sa' case are in general good agreement with these lensing and dynamics analyses.

The results of the two combined strong lensing, stellar dynamics, and stellar population studies by \citet{tre10} and \citet{car11} on samples of more than 50 SLACS lens galaxies agree on finding that a constant heavy (Salpeter-like) stellar IMF requires a shallower dark-matter density profile than a constant light (Chabrier-like) stellar IMF. Furthermore, based on different samples of SLACS lenses and mass diagnostics, \citet{jia07} and \citet{aug10b} conclude that adiabatically compressed models of the galaxy dark-matter halos are favored. These findings are also in qualitative agreement with our results.

The next natural step towards a clearer picture of the internal structure of massive early-type galaxies will be the extension of the SLACS sample to the lens galaxies selected from the BOSS (Baryon Oscillation Spectroscopic Survey; \citealt{eis11}) Emission-Line Lens Survey (BELLS; \citealt{bro12}). The lens galaxies for which strong lensing and stellar population models will be available at the end of this new survey will allow to enlarge significantly the lens sample (in nearly the same luminous mass range) and to explore the average density profile of the galaxy dark-matter halos on a radial range (i.e., $\Lambda$) that is approximately twice as large as done here. 

\section{Conclusions}

We have combined strong gravitational lensing and stellar population synthesis models in a homogeneous sample of massive early-type galaxies to measure the logarithmic inner slope of the average dark-matter density profile. We have obtained clear indication of the contraction of the halos when compared to the results of dark matter-only cosmological simulations. This is in line with the recent findings of high-resolution hydrodynamical simulations which include radiative cooling and feedback processes (e.g., \citealt{aba10}; \citealt{tis10}; \citealt{duf10}). These studies show that the contraction of a halo does depend not only on the amount and distribution of the baryonic mass condensed at the halo centre, but also on the details of the halo assembly history. Future theoretical and observational efforts towards a better understanding of the inner dark-matter structure and the stellar initial mass function of galaxies will therefore be crucial to explore different cosmological models and to investigate the nature of dark matter and its interaction with baryons.

\acknowledgments

C. G. is grateful to Marco Lombardi, Giuseppe Bertin, Matteo Barnab\`e, and Simona Vegetti for interesting discussions. This research was supported by the DFG cluster of excellence ``Origin and Structure of the Universe''. 






\clearpage




\clearpage

\clearpage

\end{document}